\documentclass[journal,twoside,web]{ieeecolor}
\usepackage{main-style}
\usepackage{cite}
\usepackage{amsmath,amssymb,amsfonts}
\usepackage{algorithmic}
\usepackage{graphicx}
\usepackage{textcomp}

\usepackage{adjustbox}
\usepackage{siunitx}
\usepackage{caption}
\usepackage[T1]{fontenc}
\usepackage{hyperref}
\usepackage{multirow}
\usepackage{bm}

\newcommand\Tstrut{\rule{0pt}{2.3ex}}         
\newcommand\TstrutL{\rule{0pt}{2.7ex}}         

\def\BibTeX{{\rm B\kern-.05em{\sc i\kern-.025em b}\kern-.08em
    T\kern-.1667em\lower.7ex\hbox{E}\kern-.125emX}}
\begin{document}

\title{Synthetic optical coherence tomography angiographs for detailed retinal vessel segmentation without human annotations}

\author{Linus Kreitner, Johannes C. Paetzold, Nikolaus Rauch, Chen Chen, Ahmed M. Hagag, Alaa E. Fayed, Sobha Sivaprasad, Sebastian Rausch, Julian Weichsel, Bjoern H. Menze, Matthias Harders, Benjamin Knier, Daniel Rueckert, and Martin J. Menten
\thanks{ L.\,K., D.\,R. and M.\,J.\,M. are affiliated with the Lab for AI in Medicine, Klinikum rechts der Isar, Technical University of Munich, Germany. J.\,C.\,P., C.\,C., D.\,R. and M.\,J.\,M are with BioMedIA, Imperial College London, United Kingdom (UK). J.\,C.\,P is also affiliated with ITERM Institute Helmholtz Zentrum Muenchen, Germany. N.\,R. and M.\,H. are with the Interactive Graphics and Simulation Group, University of Innsbruck, Austria. C.\,C. is also with the Oxford BioMedIA group, University of Oxford, UK. A.\,M.\,H. and S.\,S. are with the NIHR Moorfields Biomedical Research Centre, Moorfields Eye Hospital NHS Foundation Trust, London, UK. A.\,M.\,H. is also with Boehringer Ingelheim Limited, UK. S.\,S. is with University College London, UK. A.\,E.\,F. is with the Department of Ophthalmology, Kasr Al-Ainy School of Medicine, Cairo University, Egypt and the Watany Eye Hospital, Cairo, Egypt. S.\,R. and J.\,W. are with Heidelberg Engineering GmbH, Heidelberg, Germany. B.\,H.\,M. is with the Department of Quantitative Biomedicine, University of Zurich, Switzerland. B.\.K. is with the Department of Neurology, Klinikum rechts der Isar, Technical University of Munich, Germany.}}

\maketitle

\begin{abstract}
Optical coherence tomography angiography (OCTA) is a non-invasive imaging modality that can acquire high-resolution volumes of the retinal vasculature and aid the diagnosis of ocular, neurological and cardiac diseases. Segmenting the visible blood vessels is a common first step when extracting quantitative biomarkers from these images. Classical segmentation algorithms based on thresholding are strongly affected by image artifacts and limited signal-to-noise ratio. The use of modern, deep learning-based segmentation methods has been inhibited by a lack of large datasets with detailed annotations of the blood vessels. To address this issue, recent work has employed transfer learning, where a segmentation network is trained on synthetic OCTA images and is then applied to real data.
However, the previously proposed simulations fail to faithfully model the retinal vasculature and do not provide effective domain adaptation. Because of this, current methods are unable to fully segment the retinal vasculature, in particular the smallest capillaries.
In this work, we present a lightweight simulation of the retinal vascular network based on space colonization for faster and more realistic OCTA synthesis. We then introduce three contrast adaptation pipelines to decrease the domain gap between real and artificial images. We demonstrate the superior segmentation performance of our approach in extensive quantitative and qualitative experiments on three public datasets that compare our method to traditional computer vision algorithms and supervised training using human annotations.
Finally, we make our entire pipeline publicly available, including the source code, pretrained models, and a large dataset of synthetic OCTA images.
\end{abstract}


\begin{figure*}[htbp]
  \centering
  \includegraphics[width=2\columnwidth]{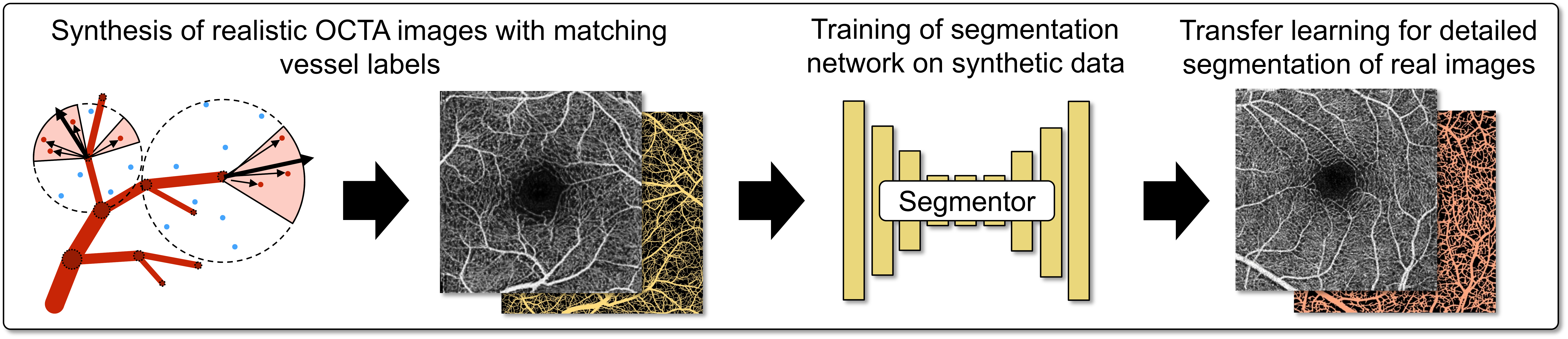}
  \caption{
  Our proposed pipeline generates realistic synthetic OCTA images and uses them to train a U-Net for blood vessel segmentation. When applied to real data, the trained network is able to produce highly detailed segmentation maps.}
  \label{figure:abstract}
\end{figure*}

\section{Introduction}
\label{introduction}

Optical coherence tomography angiography (OCTA) is a non-invasive imaging modality that can acquire high-resolution volumes of the retinal vasculature. The technology can aid the diagnosis of ocular, neurological, and cardiac diseases \cite{JacquelineChua.2020, J.Wang.2019, Aly.2022}. Recent work has investigated the feasibility of automated disease classification and grading based on OCTA images in an end-to-end approach using convolutional neural networks (CNN) \cite{Yasser.2022}. However, these systems only offer limited explainability and might be subject to unknown training biases. For this reason, others first extract a vessel segmentation map and then compute disease-relevant biomarkers, such as vessel density, radius, or tortuosity.

Currently, most vessel segmentation methods for OCTA images are based on thresholding algorithms. The most reliable choices are found to be adaptive thresholding, Frangi filters with binarization, Gabor filters with binarization, and optimally oriented flux \cite{Giarratano.2020}. However, all methods require manual tuning and suffer from poor robustness towards the diverse set of image artifacts in real OCTA data. Furthermore, small vessels such as the capillaries from the deep vascular complex (DVC) are hard to detect using thresholding but crucial for early disease detection \cite{Meiburger.2021}.

Several machine learning based vessel segmentation solutions for OCTA images have been proposed in the past. Pissas \textit{et al.} train a U-Net on 50 8$\times$8\,mm$^2$ manually annotated OCTA images that iteratively refines the segmentation map \cite{TheodorosPissas.2020}. Mou \textit{et al.} propose a generalized segmentation model using spatial and channel attention for curvilinear structures \cite{LeiMou.2021}. They demonstrate their performance on corneal confocal microscopy images, color fundus images and an in-house data collection of 30 retinal OCTA scans of the superficial vascular complex (SVC) that were manually annotated by an expert.
However, overall, the adoption of machine learning-based vessel segmentation algorithms is hindered by a lack of sufficient training labels. Manual annotation is not only time-consuming, but also difficult due to complex vessel branching, low resolution, and low-contrast areas. Under great effort, a limited number of OCTA datasets with vessel annotations have been published  \cite{Ma.2021,Li.14.12.2020,Giarratano.2020}. However, they mostly do not contain labels for smaller vessels and sometimes exhibit labeling inconsistencies.

In settings without a sufficient amount of labeled training data, transfer learning can be used. Thereby, a network is trained in a related data domain, in which a large amount of annotations is available, before being applied to the target domain.
Costa \textit{et al.} propose generating synthetic training samples that can be used to train a network for fundus vessel segmentation \cite{Costa.2018}. An autoencoder network generates vessel graphs following the data distribution of an annotated set of images. In a second step, a generative adversarial network (GAN) transforms the vessel graphs into realistic looking fundus images. The authors are able to generate fairly realistic data pairs, but breaks in generated vessels and limited realism of the images inhibit the downstream segmentation performance on real data.
Menti \textit{et al.} use an active shape model approach to generate retinal vessel structures within a distribution observed from annotated fundus images \cite{Menti.2016}. To simulate the look of real fundus images, the authors employ handcrafted filters to augment the synthetic vessel maps.
Fu \textit{et al.} propose to train a U-Net for image denoising on paired fundus images and apply it on OCTA \textit{en-face} images for vessel enhancement \cite{MaierHein.2022}. Their transfer learning method surpasses the segmentation performance of supervised methods on a small evaluation dataset.
Several works have explored the generation of realistic medical images based on sketch maps, which are simplified binary drawings of scenes or objects \cite{JiaminLiang.2022,Wang.2023,HeZhao.2018,Zhang.2019}. For instance, Zhang \textit{et al.} use Sobel edge filters to automatically extract sketches from fundus images and train a GAN to recreate the images \cite{Zhang.2019}. A second GAN learns to generate new synthetic sketches. These synthetic sketch-image pairs are then used to pretrain a vessel detection network. Although the sketch maps do not correspond to the true vessel segmentations, the pretraining boosts the performance on a small evaluation dataset.
All the mentioned works either still require at least a small set of labeled data or only segment vessels to a limited level of detail. Furthermore, none of the methods have been tested on a larger set of OCTA images.

Recently, Menten \textit{et al.} presented a novel approach for OCTA data that does not require any labeled data \cite{Menten.2022}. They leverage a physiology-based simulation model to generate artificial vessel maps of the retinal vasculature with inherently matching labels. A CNN is then trained on these synthetic OCTA images in a supervised fashion (see figure \ref{figure:abstract}). Several image augmentations simulate OCTA artifacts and make the network more robust against the domain shift when the pre-trained network is applied to real data.
However, their work has several limitations: First, the rigid nature of the simulation model together with image synthesis taking up to eight hours per sample prevents the simulation of smaller capillaries in the retinal vasculature. It further complicates hyperparameter tuning to increase the image realism. Second, the manually tuned image augmentations are not able to fully bridge the domain gap between synthetic and real images. Because of this, the method proposed by Menten \textit{et al.} is not able to fully segment real OCTA images, partially ignoring the smallest capillaries. Third, Menten \textit{et al.} only compare their method to a single dataset and do not explore the clinically important task of segmenting vessels beyond the level of detail of existing annotations.

\begin{figure*}[htbp]
  \centering
  \includegraphics[width=1.6\columnwidth]{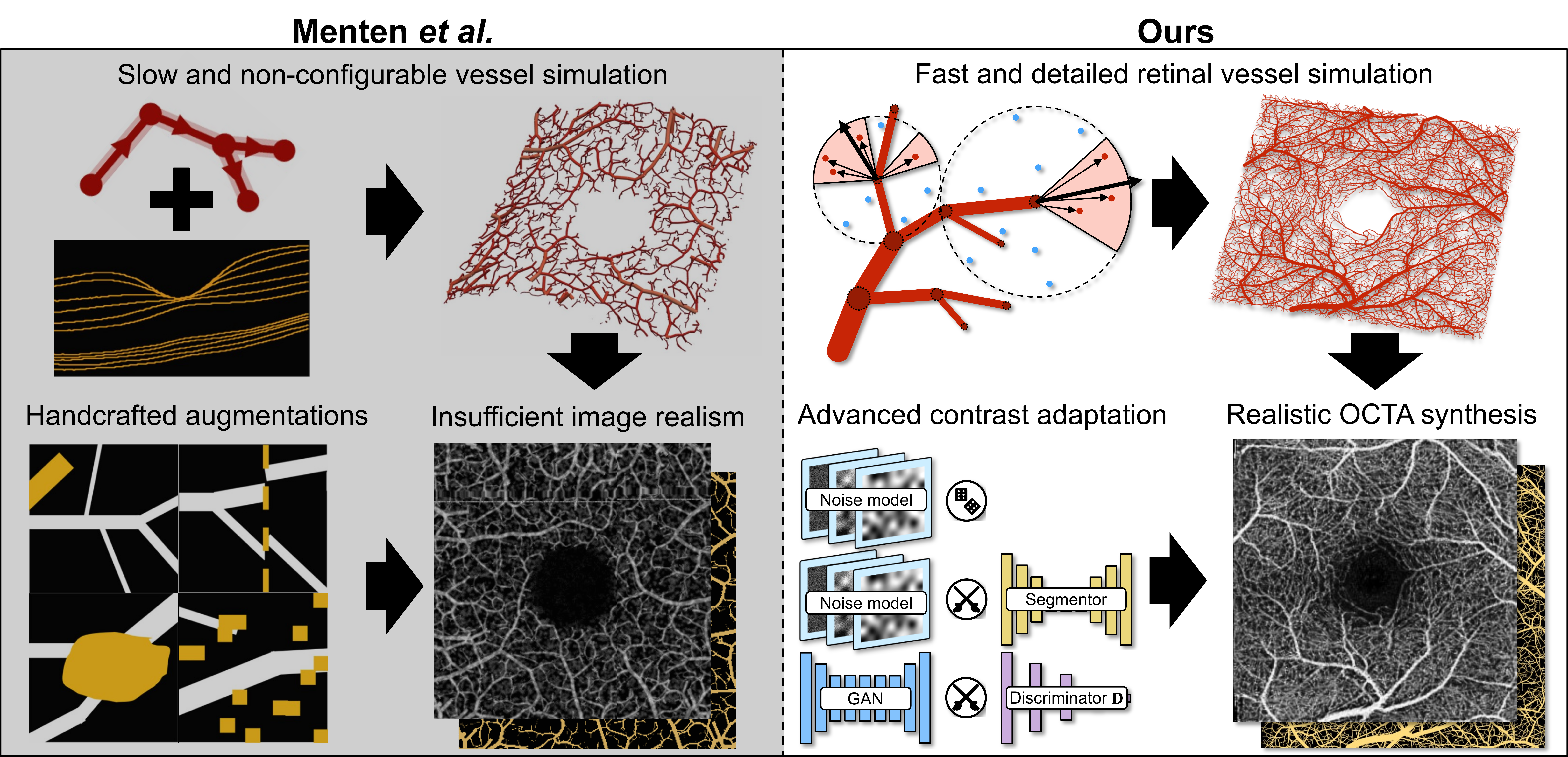}
  \caption{
  Comparison of Menten \textit{et al.}'s image synthesis and ours. Synthetic images are generated by 1) growing a vascular network via an angiogenesis driven simulation model, and 2) enhancing vessel map realism with data augmentations. Our new simulation model enables faster synthesis of complex vessel structures, and our data driven contrast adaption strategies model artifacts more realistically. Figures of the method by Menten \textit{et al.} are taken from \cite{Menten.2022}.
  }
  \label{figure:abstract2}
\end{figure*}

In this work, we introduce a pipeline to train a CNN for vessel segmentation of 2D macular OCTA images using synthetic OCTA images that overcomes the aforementioned disadvantages. To this end, we make the following key contributions:
\begin{enumerate}
    \item We replace the computationally costly physiology-based simulation with a statistical angiogenesis model based on space colonization from Rauch and Harders \cite{Rauch.2021} (see figure \ref{figure:abstract2}). The new algorithm substantially speeds up the image generation and enables us to quickly test new configurations. With our method, we can control the look and dynamics of the vessel growth, leading to a more complex and realistic vasculature of the retina.
    \item We employ and compare several advanced contrast adaptation strategies to bridge the domain gap between real and synthetic images. We show that this step is crucial for the segmentation robustness on real data.
    \item We perform extensive quantitative and qualitative benchmarking on three public datasets. We compare our methods with traditional computer vision algorithms and models trained directly on the dataset labels.
    \item We published our entire pipeline as an open source tool, including a synthetic dataset and pretrained models at \url{https://github.com/TUM-AIMED/OCTA-seg}.
\end{enumerate}

\section{Methods}
\subsection{Statistical simulation of the retinal vasculature}
A central component of our pipeline is the realistic simulation of retinal vasculature. Originating from the optical nerve, blood vessels traverse the entire retina as two vascular complexes. The superficial vascular complex (SVC) mostly contains larger vessels, while the deep vascular complex (DVC) consists primarily of capillaries with a radius as small as 2.5$\,\mu$m \cite{Campbell.2017}. The retina also contains a circular shaped avascular zone around the fovea (FAZ) of about 0.35\,mm in diameter. Similar to other works, we simulate vessel development using a forest of 3D rooted binary trees \cite{Rauch.2021, Schneider.2012, Nekka.1996, Talou.2021}. Growth of these graphs is governed by a set of predefined rules. Each tree is initialized by a root node with a single child node. An edge encodes the length and radius of a vessel segment. There are three types of nodes:
\begin{itemize}
    \item \textbf{Leaf-node:} The node does not have any children and is only connected to its parent.
    \item \textbf{Inter-node:} The node has exactly one child.
    \item \textbf{Bounded node:} The node has reached its maximum of two children and is not considered for proliferation.
\end{itemize}
In clinical practice, OCTA images are mostly viewed as 2D \textit{en-face} projection representation instead of the full 3D volume.
We therefore relax the realism in depth, as we are only interested in the resulting 2D maximum intensity projection (MIP) along the z-axis. To simulate the geometrical shape of a 3$\times$3\,mm$^2$ central crop of the retina, we define a simulation space with dimensions of 3$\times$3$\times \frac{1}{76}$\,mm$^3$. The dimensions of the simulation space can also be scaled to support different fields of view (FOVs). Since the optical nerve is located outside the FOV for macular images, we randomly place 16 root stumps at the lateral faces of the simulation space cuboid. We note that it is also possible to define a circular zone inside the simulation space to replicate the optical nerve for ultra-wide field OCTA images.

Menten \textit{et al.} base their growth model on an angiogenesis simulation by Schneider \textit{et al.} \cite{Schneider.2012}, which has been frequently used to create synthetic blood vessels \cite{gerl2020distance, todorov2020machine}. Inspired by angiogenesis in biological organisms, they model the diffusion of oxygen (O$_2$) and the related vascular endothelial growth factor concentration iteratively for the entire volume to control the sprouting of new vessels from leaf- or inter-nodes. The repeated calculation of the oxygenation for each vessel sprout is computationally very expensive and leads to an excessive synthesis duration of about eight hours per image for Menten \textit{et al.} This prevents the generation of more complex vessel graphs and complicates further development of the algorithm. Additionally, the complexity of the simulation leads to a reduced configurability and inhibits the generation of vessel trees with specific growth pattern. 
Rauch and Harders present a simplified angiogenesis model that completely avoids the explicit calculation of the oxygen concentration by using a statistical approach based on space colonization \cite{Rauch.2021}. In this work, we adapt this approach and will discuss the algorithm in the following. A list of all mentioned variables is provided in table \ref{table:params}.

\begin{table}[htbp]
\centering
\caption{The parameters used by our angiogenesis model. The growth process of the OCTA volume is divided into a phase to simulate the SVC, and the second phase to generate the smaller and more dense vessels of the DVC.}
\label{table:params}
\begin{tabular}{llcc}
\hline
\Tstrut
Var & Description & SVC & DVC\\ \hline
\Tstrut
\underline{r$_\textnormal{faz}$} &
  Radius of the FAZ [mm] &
  \multicolumn{2}{c}{$\sim \mathcal{N}$(0.45,0.021)}\\ 
\underline{r$_\textnormal{rot}$} &
  Rotation effect radius around the FAZ [mm] &
  1.05 & 1.05 \\ 
I &
  Number of iterations &
  100 &
  100\\ 
N &
  Number of added OSs per iteration &
  1000 &
  2000\\ 
\underline{d} &
  Terminal segment length [mm] &
  0.1 &
  0.04\\ 
\underline{r} &
  Terminal vessel radius [mm] &
  0.0025 &
  0.0025\\
\underline{$\epsilon_n$} &
  Min. distance between nodes and OSs [mm] &
  \multicolumn{2}{c}{See equation \ref{equation-eps_n}} \\
\underline{$\epsilon_s$} &
  Min. distance between OSs [mm] &
  0.135 &
  0.045\\
\underline{$\epsilon_k$} &
  Satisfaction range of nodes [mm] &
  0.135 &
  0.045 \\
\underline{$\delta$} &
  Distance of perception volume [mm] &
  0.2925 &
  0.0975 \\
$\gamma$ &
  Angle of perception cone w.r.t. parent &
  50° &
  90° \\
$\phi$ &
  Min. std. of attr. vector angles for sym. bif. &
  15° &
  15°\\
$\omega$ &
  Weighting factor optimal growth direction &
  0.3 &
  0 \\
$\kappa$ &
  Bifurcation exponent &
  2.55 &
  2.9 \\
$\Delta \sigma$ &
  Linear growth rate of sim. space per iter. &
  0.02 &
  0.02 \\ \hline
\end{tabular}
\end{table}

\subsubsection{Oxygen sink placement}
Rauch and Harders model the circulatory system with oxygen-providing arterial trees and CO$_2$-removing venous trees \cite{Rauch.2021}. Instead of explicitly computing the oxygen distribution, the authors propose to randomly place $N$ oxygen sinks (OSs) in the simulation space every iteration. Each OS acts as an attraction point for its closest arterial node within range $\delta$. Each OS must have a minimal distance $\epsilon_s$ to existing OSs and a minimal distance $\epsilon_n$ to nearby arterial nodes. In reality, smaller vessels supply more oxygen to surrounding tissue than larger ones. We model this by allowing new OSs to be placed closer to arterial nodes as the vessels' diameter increase. We use the oxygen concentration heuristic defined by Schneider \textit{et al.} \cite{Schneider.2012}, and set $\epsilon_n$ relative to a vessel's radius:
\begin{equation}
\label{equation-eps_n}
    \epsilon_n = 0.02\cdot203.9\,\frac{\mu \textnormal{l}}{\textnormal{ml}} \cdot \frac{r_\textnormal{node}}{3.5\,\mu\textnormal{m}}\exp\left(1-\frac{r_\textnormal{node}}{3.5\,\mu\textnormal{m}}\right).
\end{equation}
Once a new vessel node is placed within range $\epsilon_k$ of an attraction point, the attraction point is considered saturated and is turned into a CO$_2$ source. These CO$_2$ emitters now act as attraction points for venous trees, whose growth process follows analogue to arterial trees. Once CO$_2$ sources are satisfied, they are removed.

\subsubsection{Leaf-node proliferation}
\begin{figure}[htbp]
  \centering
  \includegraphics[width=\columnwidth]{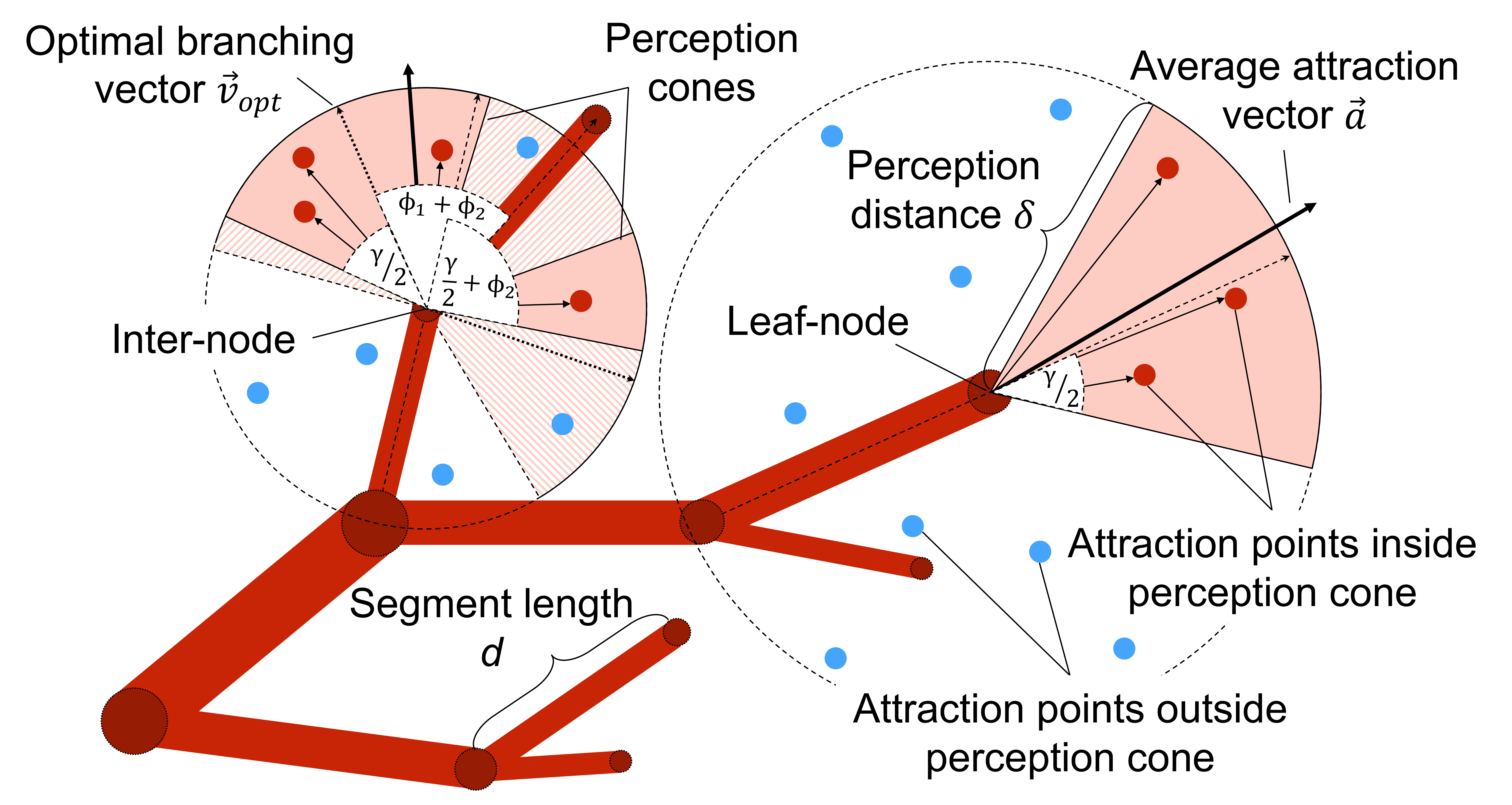}
  \caption{Perception volumes for inter-node sprouting (left) and elongation (right) in 2D. The growth direction for the new node is given by a combination of the mean attraction vector and the optimal branching vector.}
  \label{figure:vessel-growth}
\end{figure}

Every leaf node has a perception cone with angle $\gamma$ and distance $\delta$ (see figure \ref{figure:vessel-growth}). A given attraction point at position $\vec{p}_{\textnormal{att}}$ is considered by the node for proliferation if:
\begin{equation}
    \left\Vert \vec{p}_{\textnormal{node}} - \vec{p}_{\textnormal{att}} \right\Vert_2 \leq \delta 
    \quad \mathrm{and} \quad
    \angle \vec{p}_{\textnormal{parent}}\vec{p}_{\textnormal{node}}\vec{p}_{\textnormal{att}} \leq \frac{\gamma}{2}.
\end{equation}
Let $S_{\textnormal{att}}$ be the set of considered attraction points and $\vec{v}_\textnormal{opt}$ the optimal branching unit vector. For leaf-nodes, $\vec{v}_\textnormal{opt}$ denotes the vector of the parent to the current node. A new child node is placed in the direction of the elongation vector $\vec{g}$ at position $\vec{p}_{\textnormal{new}}$:
\begin{equation}
\label{equation:node-placement}
    \begin{aligned}
     &\vec{g} =\textnormal{norm}\left( \omega \vec{v}_\textnormal{opt} + (1-\omega)\sum_{\vec{p}_{\textnormal{att}} \in S_\textnormal{att} } \textnormal{norm}(\vec{p}_{\textnormal{att}} - \vec{p}_{\textnormal{node}} ) \right)\\
    &\vec{p}_{\textnormal{new}} = \vec{p}_{\textnormal{node}} + d \vec{g}
    \end{aligned}
\end{equation}
norm($\cdot$) normalizes the vector by its L$_2$ norm and $\omega$ is a weighting factor to control the allowed deviation of the average attraction vector $\vec{a}$ to the hypothetical optimal branching vector $\vec{v}_\textnormal{opt}$. $d$ is the fixed length of the segment (see section \ref{sec:simspace-expansion}). The radius of the new segment is set to the fixed terminal vessel radius $r$. This process is called \textit{elongation}. If the angle of all attraction vectors is larger than a threshold $\phi$, a \textit{bifurcation} is initiated instead and two child nodes are added. We set the radii $r_{c1}$ and $r_{c2}$ of the two child nodes to the terminal vessel radius $r$. The angles $\alpha$ and $\beta$ from $\vec{a}$ to the child segments are calculated following Murray's law of minimum work \cite{Murray.1926}.
\begin{equation}
    \alpha = \cos^{-1}\left(\frac{r^4_{\textnormal{node}}}{ 2r^2_{\textnormal{node}}r^2}\right) = -\beta.
\end{equation}
The parent vessel segment radius $r_{\textnormal{node}}$ is updated to satisfy
\begin{equation}
    r^\kappa_{\textnormal{node}} = r^\kappa_{\textnormal{c1}} + r^\kappa_{\textnormal{c2}},
\end{equation}
where $\kappa$ denotes the bifurcation exponent. This radius adjustment is recursively repeated up to the root node. The child nodes are placed in the plane spanned by the leaf node and the line that cuts through the mean of all attraction points while minimizing the orthogonal distance to them.

\subsubsection{Inter-node sprouting}
Attraction points are considered by an inter-node if they lay within the frustum of a sphere with radius $\delta$,
\begin{equation}
    \begin{aligned}
        \phi_1 + \phi_2 - \frac{\gamma}{2} \leq \ &\angle \vec{p}_{\textnormal{node}}\vec{p}_{\textnormal{child}}\vec{p}_{\textnormal{att}} \leq \phi_1 + \phi_2 + \frac{\gamma}{2},\\
         \textnormal{and} \quad &\angle \vec{p}_{\textnormal{parent}}\vec{p}_{\textnormal{node}}\vec{p}_{\textnormal{att}} \leq \frac{\gamma}{2} + \phi_2 .
    \end{aligned}
\end{equation}
Here we extend Rauch and Harders' definition of the perception volume to prevent unrealistic branching angles. To find the optimal growth direction, we look at the set $S_{\vec{v}_{\textnormal{opt}}}$ of hypothetical optimal branching vectors towards the existing child segment following Murray's law. The vectors can be thought of as freely rotating around the child segment at an angle $\phi_2$. The closest optimal branching vector $\vec{v}_{\textnormal{opt}} \in S_{\vec{v}_{\textnormal{opt}}}$ to the average attraction vector $\vec{a}$ is used for proliferation.
The new child node's position is computed following equation \ref{equation:node-placement}.

\subsubsection{Simulation space expansion}
\label{sec:simspace-expansion}
To ensure that the vasculature network grows homogeneously in the entire simulation space while not being limited by its size, we expand the simulation space as the networks grows. For this, all distance related parameters are linearly reduced every iteration. Specifically, a parameter $p^{(t)}$ at time step $t$ is given by dividing the initial value $p^{(0)}$ with the scaling factor $\sigma^{(t)} = 1+t\times \Delta \sigma$.
We keep the terminal vessel radius $r$ fixed to the minimal vessel size, and only increase a vessel's diameter following Murray's law. This prevents the placement of abnormally large vessel stumps. Additionally, we shrink the segment length parameter $d$ only to a minimal value of 0.04$\,$mm.
As listed in table \ref{table:params}, the SVC and the DVC are grown using two different sets of parameters. Due to this parameter adjustment, the vessel trees of the DVC exhibit the formation of smaller and more dense branches. Given that the ultimate image is generated through the MIP along the depth dimension, we allow both complexes to develop in the same volume. Optionally, it is also possible to grow the vessel graphs separately to simulate each complex in isolation.

\subsubsection{Simulation of the foveal avascular zone}

\begin{figure}[htbp]
  \centering
 \includegraphics[width=\columnwidth]{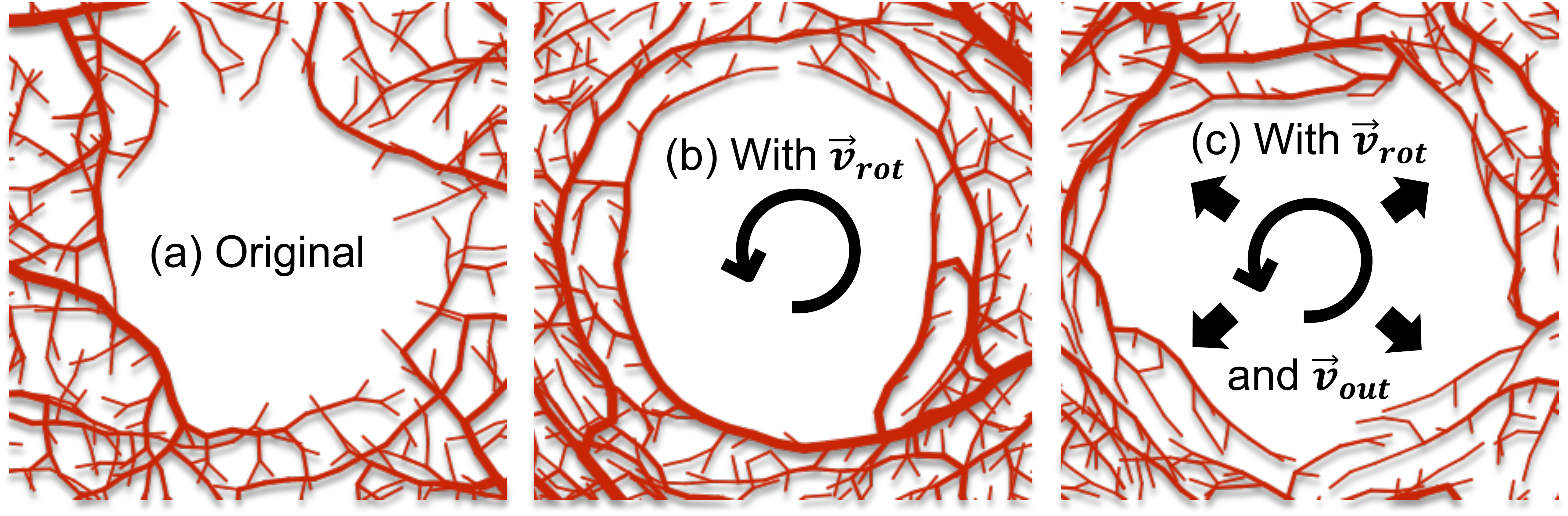}
  \caption{Adding a rotation vector $\vec{v}_{\textnormal{rot}}$ and an out vector $\vec{v}_{\textnormal{out}}$ to the elongation vector leads to a more realistic growth pattern around the FAZ.}
  \label{figure:faz}
\end{figure}

To recreate the shape of the FAZ, we extend the algorithm with several rules. To prevent vessel growth inside the FAZ, we avoid the placement of OSs within a central circular region with radius $r_{\textnormal{faz}}$. As the retinal vasculature features fewer bifurcations close to the FAZ, we decrease the chance of bifurcations and inter-node sprouting in regions closer to the center of the FAZ, and instead perform a simple elongation for leaf nodes.
To achieve a more circular look of the FAZ, we add a rotation term to the elongation vector $\vec{g}$. We calculate the orthogonal rotation vector $\vec{v}_{rot}$ by a $\ang{90}$ rotation of $\vec{c} = \vec{p}_{\textnormal{center}} - \vec{p}_{\textnormal{node}}$ towards the direction of the average attraction vector $\vec{a}$. To prevent a degenerate fully circular growth pattern, we also add a weighted term $\vec{v}_{\textnormal{out}}$ to the elongation vector that gradually motivates growth away from the center. The vector is given by the negative center vector $\vec{c}$. Let $w = \sqrt{r_{\textnormal{rot}} - \left\Vert \vec{c}\right\Vert_2}$ be the distance-dependent weighting factor. The position for a new node after elongation is given by
\begin{equation}
    \vec{p}_{\textnormal{new}} = \vec{p}_{\textnormal{node}} + d \cdot \textnormal{norm}\left( (1-w)\vec{g} + \frac{2w}{3}\vec{v}_{\textnormal{rot}} +\frac{w}{3}\vec{v}_{\textnormal{out} }\right).
\end{equation}
In reality, we observe varying FAZ shapes and radii. We therefore randomly choose $r_{\textnormal{faz}}$ from a normal distribution. The effects of these additions can be seen in figure \ref{figure:faz}.

\subsection{Image and label generation from vascular graphs}
After generating the vessel graphs, we voxelize the simulation space and take the maximum intensity projection along the z-dimension to compute the anti-aliased grayscale images.
The synthetic vessel annotations $bin(\widetilde{X})$ are given by binarizing the images. We note that segmentation maps necessitate a specific resolution in order to accurately distinguish individual small blood vessels. We therefore upsample all OCTA images via bilinear interpolation $bilin(\cdot)$ to a pixel size equal to the minimal vessel radius of approximately 2.5$\,\mu$m. For instance, 3$\times$3$\,$mm$^2$ images with a dimension of 304$\times$304 pixels are upsampled to 1216$\times$1216 pixels. The synthetic segmentation labels are directly generated in the upsampled simulation space to ensure correct vessel diameter representation. Table \ref{table:namings} provides an overview over naming conventions we use to distinguish the images.

\begin{table*}[htbp]
\centering
\caption{Naming conventions for inputs, intermediates, and outputs of our framework. LR stands for low resolution ($1\,\textnormal{pixel}\ \widehat{\approx}\ $10$\times$10$\,\mu\textnormal{m}^2$), and HR for high resolution ($1\,\textnormal{pixel}\ \widehat{\approx}\ $2.5$\times$2.5$\,\mu\textnormal{m}^2$).}
\label{table:namings}
\begin{tabular*}{\linewidth}{@{\extracolsep{\fill}}clcl}
\hline
\Tstrut
Var & Name & Resolution & Description\\ \hline
\Tstrut
    $VG$ & Vessel graph & continuous & Forest of binary trees in 3D simulation space generated by our simulation model \\
    $X$ & Synthetic vessel map & LR & Rasterized grayscale image of $VG$ \\
    $\widetilde{X}$ & Synthetic vessel map & HR & Rasterized grayscale image of $VG$ \\
    $bilin(\cdot)$ & Upsampled image & HR & Upsampled image via bilinear interpolation \\
    $G(X)$ & Artificial OCTA image & LR & $X$ with additional contrast adaptation by GAN \\
    \textit{NM}$(X)$ & Artificial OCTA image & LR & $X$ with additional contrast adaptation by handcrafted noise model \\
    $bin(\widetilde{X})$ & Synthetic annotations & HR & Vessel segmentation mask from binarized synthetic vessel map \\
    $S(\cdot)$ & Predicted segmentation & HR & Binary segmentation mask predicted by segmentor $S$ \\
    $Y$ & Real OCTA image & LR & Real human OCTA image from public dataset \\
\hline
\end{tabular*}
\end{table*}

\begin{figure*}[htbp]
  \centering
  \includegraphics[width=1.9\columnwidth]{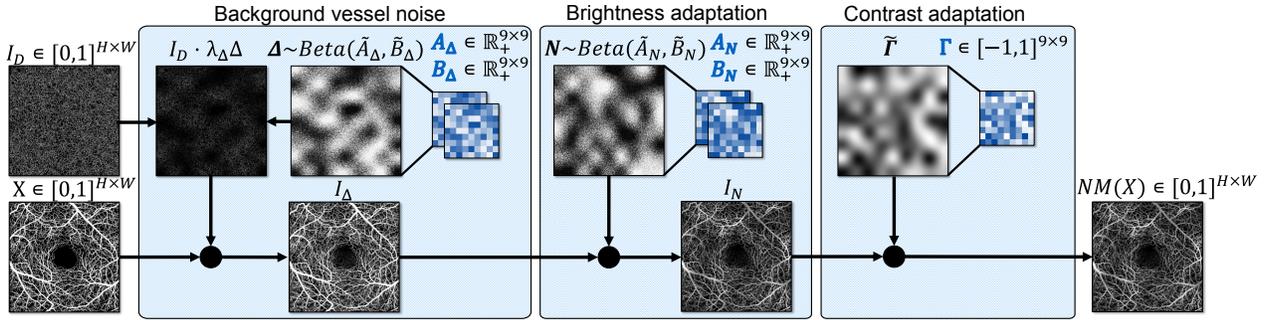}
  \caption{Structure of our proposed handcrafted noise model to simulate artifacts and contrast variations of real OCTA images. Given a synthetic vessel map $X$ and a background vessel map $I_D$, the module successively performs 1) background noise addition, 2) brightness augmentation, and 3) contrast adaptation. In each block, we use a sparse control point matrix to generate a field for locally varying contrast.
  }
  \label{figure:noise_model}
\end{figure*}

\subsection{Domain adaptation of synthetic images}
\label{sec:domain-adaptation}
Using the synthetic vessel maps $X$ directly to train a segmentation network results in poor performance on real data. Although we realistically simulate the vessel geometry, the contrast and SNR of the images are different. In practice, OCTA images exhibit artifacts such as noise, low-contrast regions, and blurry edges. Menten \textit{et al.} apply shearing, stretching, and binomial noise artifacts on their training corpus. However, they find that these augmentations do not substantially boost their segmentation performance. In the following, we introduce three advanced strategies to adapt the contrast of training images with the ultimate goal of improving the segmentation performance on real images.

\subsubsection{Handcrafted noise model}
\label{sec:hand-noise}
Our first approach is a handcrafted multistep noise model designed to recreate the artifacts observed in real data (see figure \ref{figure:noise_model}). First, we model the structured background noise that is caused by sub-resolution capillaries in the image. We use our vascular simulation to generate a dense background noise image $I_D\in [0,1]^{H \times W}$ of capillaries. We multiply this layer with an additional Beta distribution noise layer $\Delta$ to locally modulate the SNR. Depending on the parameters of the Beta distribution, the noise can exhibit properties of a uniform, normal, or even binary distribution.
Instead of a single noise distribution for the entire image, we use a sparse grid of control points that govern the noise in the area surrounding them \cite{Chen.2020}. Let $A_\Delta \in \mathbb{R}_+^{9\times9}$ and $B_\Delta \in \mathbb{R}_+^{9\times9}$ be the control points for the parameters $\alpha$ and $\beta$ of the beta distribution. We use bicubic interpolation to compute the pixelwise parameters $\widetilde{A}_\Delta$ and $\widetilde{B}_\Delta$. We then sample the modulation factor for each pixel as $\Delta^{(i)} \sim \textnormal{Beta}(\widetilde{A}^{(i)}_\Delta, \widetilde{B}^{(i)}_\Delta)$.

Second, we use the same procedure to generate a Beta noise matrix $N$ from a set of control points to model speckle noise and local brightness adjustments.
Third, we generate a field for locally varying contrast. Let $\Gamma \in [-1, 1]^{9\times9}$ 
denote the contrast control points and $\widetilde{\Gamma} = \textnormal{Bicubic}(\Gamma)$ the resulting matrix after interpolation. 

Each transformation step is parameterized by a weighting factor $\lambda$. The final output \textit{NM}$(X)$ of the noise model is given by:
\begin{equation}
    \begin{aligned}
        I_\Delta &= \textnormal{MAX}[X, I_D \cdot \lambda_\Delta\Delta]\\
        I_N &= (1-\lambda_N)I_\Delta + \lambda_NNI_\Delta\\
        NM(X) &= (I_N)^{(\lambda_\Gamma\widetilde{\Gamma}+1)}.
    \end{aligned}
\end{equation}
To simulate blurry edges and lower quality images, we apply random down-sampling by a factor $s\sim U(0.25,1)$ and a successive up-sampling to the original size.

\subsubsection{Noise modeling via adversarial training}
\begin{figure*}[htbp]
  \centering
  \includegraphics[width=1.9\columnwidth]{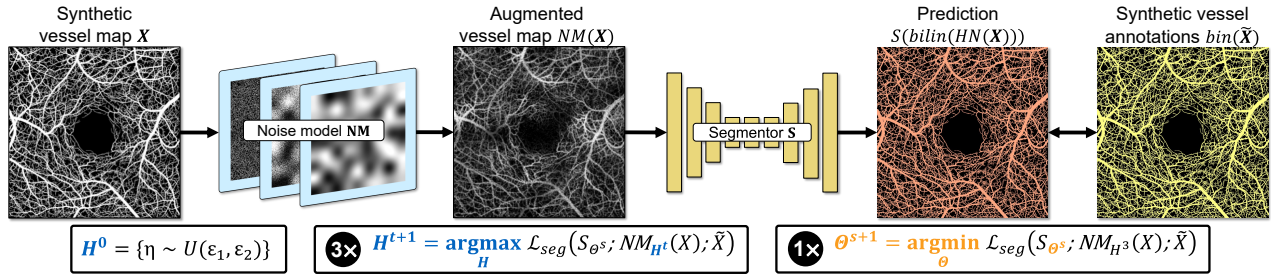}
  \caption{Our adversarial training approach automatically selects the optimal parameters of our handcrafted model in order to maximize the segmentation loss. After three projected gradient ascent optimization steps ($\mathrm{argmax}_H$), the weights of the segmentation network are updated via stochastic gradient descent ($\mathrm{argmin}_\Theta$) to minimize the loss given the adversarial sample.}
  \label{figure:adv-training}
\end{figure*}

The contrast adaptation of synthetic training images aims to ultimately improve the performance of the segmentation network. The random selection of control point values in the previously described approach does not prioritize difficult augmentations and is inefficient. To this end, we explore an adversarial training (AT) approach, where an intensity transformation is computed for each image that maximizes the segmentation loss given the training label. To prevent pixelwise attacks that could alter the alignment of images and labels, we adopt a strategy proposed by Chen \textit{et al.} \cite{Chen.2020}. In order to segment MR images in a scarce data environment, their work focuses on augmenting training samples by applying adversarially tuned magnetic resonance bias fields. They propose to only tune a small set of control points via AT that are then used to compute the transform. Integrating this into our noise model, we turn the control point values of each layer into tunable parameters and optimize them via iterative projected gradient ascent (PGA) (see figure \ref{figure:adv-training}).
Since sampling from a Beta noise distribution is not differentiable, we use Pytorch's implementation of pathwise derivatives to stochastically obtain the gradients for the $\alpha$ and $\beta$ parameters of the Beta distributions \cite{pmlr-v80-jankowiak18a}. Let $\Pi_{[\epsilon_1, \epsilon_2]}$ denote the projection of a parameter to its allowed interval, then each parameter $\eta$ of \textit{NM} is updated by:
\begin{equation}
    \eta_{\textnormal{PGA}}^{t+1} = \Pi_{[\epsilon_1, \epsilon_2]}( \eta_{\textnormal{PGA}}^{t} + lr \cdot \nabla_\eta \mathcal{L}_\textnormal{seg}(S;NM(X);\widetilde{X}))
\end{equation}

\subsubsection{Generative adversarial training}

\begin{figure*}[htbp]
  \centering
  \includegraphics[width=1.9\columnwidth]{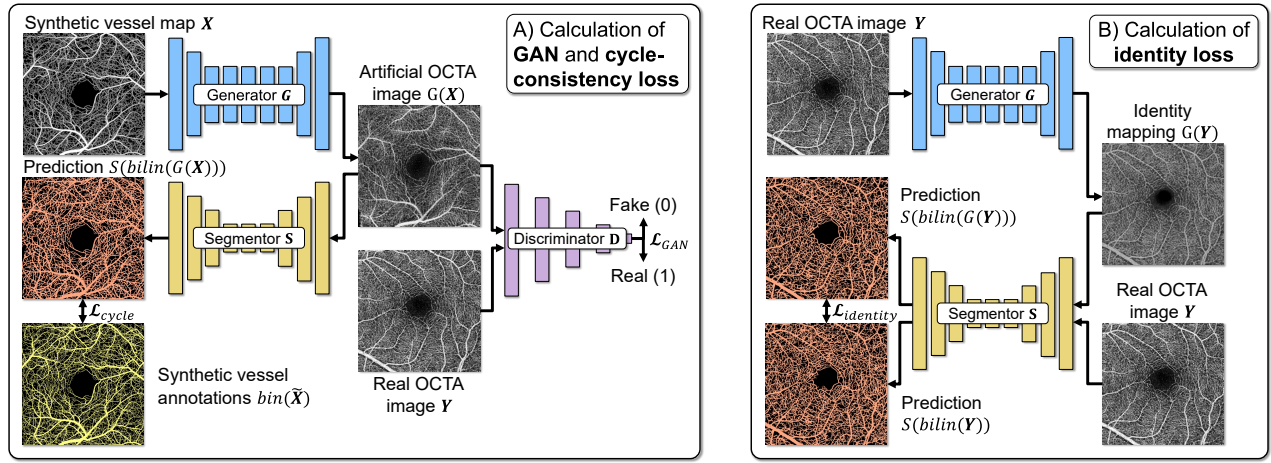}
  \caption{Our proposed framework uses a generator network to transform synthetic vessel maps into realistic OCTA images. An adversarial discriminator network is trained to differentiate real images from generated ones. A segmentation network is trained to predict the vessel segmentation from the generated image. }
  \label{figure:gan-seg}
\end{figure*}

\begin{figure}[htbp]
  \centering
  \includegraphics[width=0.95\columnwidth]{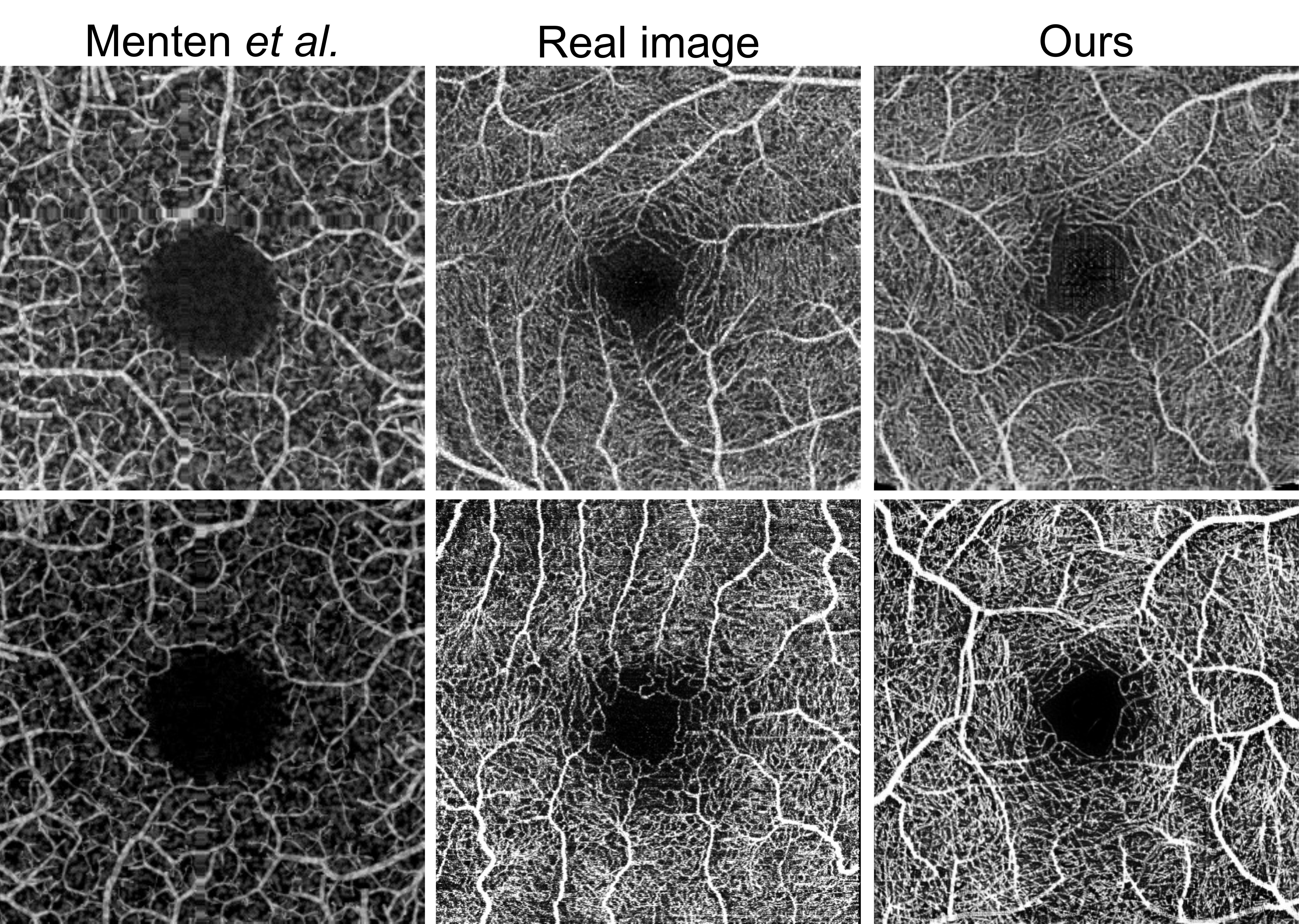}
  \caption{Our OCTA image synthesis pipeline using GAN augmentation is able to produce highly realistic training samples that closely match the geometry of the retinal vasculature and contrast of real OCTA images.}
  \label{figure:synth-comparison}
\end{figure}

Our third strategy for overcoming the domain gap is to replace the heuristically designed noise model with a learned transformation by a neural network. We formulate this task as an image-to-image style translation problem, where a generator network transforms a synthetic image from our simulation model to match the distribution and style of real images. An adversarial discriminator network judges the realism of the produced translation.

Since the generator is unconstrained in how much it transforms the image, there is no guarantee that the original image-to-label alignment is preserved. To this end, Zhu \textit{et al.} propose the CycleGAN framework to enable unpaired image-to-image translation \cite{Zhu.2017}. A second pair of generator and discriminator is used to transform the generated image back to source domain. A cycle-consistency loss term computes the difference between the input and the recovered image, encouraging the preservation of image content. CycleGAN has shown remarkable results in domain adaptation and has been used widely in medical settings \cite{Hammami.uuuuuuuu, Huang.2018, Sandfort.2019}.

In our approach, we replace the reverse generator with a segmentation network. The cycle consistency loss is directly computed using the segmentation maps that are obtained from the vascular graphs (see figure \ref{figure:gan-seg}). 
We find that this loss function is a strong regularizer for the generator and that we do not require a second discriminator. As commonly used in image-to-image translation, we use a ResNet generator with 9 residual blocks and a 70$\times$70 PatchGAN discriminator \cite{Zhu.2017, Park.2020}.
To further constrain the generator network, and to enable the segmentation network to handle real data, an identity loss term is added. To this end, the segmentation network predicts the vessel map of real images after being processed by the generator. We compare this output with the predicted segmentation of the original real image and penalize differences. Note that our approach does not require any ground truth vessel annotations of real images. To save computation time, we perform the style transfer step of the generator on the original image size and only upsample the images before segmentation. Furthermore, non-healthy samples in real data sometimes exhibit low- or non-perfusion areas. To model these, we randomly remove edges and all their descendants with probability $U(0,0.02)$ from the source-label pair. 
In future work, the basic GAN architecture could be further refined by more elaborate approaches to optimize the performance for clinical use.

\section{Experiments}
The ultimate aim of our work is to generate synthetic OCTA training images that can be used to train neural networks for segmentation of blood vessels in real OCTA images without human annotations. We now describe the evaluation procedure to compare the segmentation performance of our method with state-of-the-art baseline methods on multiple OCTA datasets.

\subsection{OCTA datasets}
We use three public datasets containing OCTA images and matching blood vessel annotations (see table \ref{table:dataset}). For all datasets, we only consider the \textit{en-face} projections of macular OCTA images. We exclusively work with images of 3$\times$3\,mm$^2$ and 304$\times$304 pixels, as they cover a sufficiently large area for analysis while maintaining a high image resolution.

\begin{table}[htbp]
\centering
\caption{Summary of all three evaluation datasets: OCTA-500 focuses on major blood vessels, while ROSE-1 provides more detailed annotations. ROSE-1, however, includes small and faint vessels inconsistently and may ignore larger ones. In some low-contrast regions, no vessels are segmented at all. Giarratano \textit{et al.}'s dataset offers the most detailed annotations.}
\label{table:dataset}
\begin{tabular}{lccc}
\hline
\TstrutL
\textbf{Name} & \textbf{Dimensions} & \textbf{No. of samples} & \textbf{Label detail} \\[1.5pt] \cline{1-4}
 \Tstrut
 OCTA-500 \cite{Li.14.12.2020} & 304$\times$304 & 200 & coarse \\
 \Tstrut
 ROSE-1 \cite{Ma.2021} & 304$\times$304 & 40 & medium \\
 \Tstrut
 Giarratano \textit{et al.} \cite{Giarratano.2020} & 91$\times$91 & 55 & detailed \\\hline
\end{tabular}
\end{table}

\subsection{Alignment of synthetic and provided vessel labels}
\label{sec:alignment}
The level of detail in the segmentation labels varies across the available datasets, which presents a challenge. This is because any segmentation predictions that surpass the level of detail provided in the human annotations would be penalized during quantitative evaluation.
To correctly evaluate the performance of a model on each dataset, we need to condition the model to produce a prediction matching the required level of detail. We observe that the datasets' segmentation labels roughly correlate with the vessel diameter. We therefore filter our synthetic segmentation labels by vessel radius to match the individual dataset labels. For instance, for the OCTA-500 dataset, we only include vessels with $r>10\,\mu$m in the segmentation map. We estimate the thresholds by manually analyzing a few samples from each set.

We note that in most use cases, we would likely aim for maximum level of detail and do not require the filtering step. However, by evaluating our generated segmentation maps with those annotated by human experts, we provide comparability to existing methods and a quantitative lower bound quality assurance of our method on multiple levels of coarseness.

\subsection{Tested segmentation methods}
We first benchmark a neural network trained directly on the small set of annotated real images and two established computer vision algorithms for vessel segmentation. We then compare these baselines with the transfer learning approach of training a network on synthetic OCTA images using the synthesis procedure proposed by Menten \textit{et al.}, and training a network on our images. We evaluate four variants of our method, to assess the performance gains of using each of our three proposed contrast adaptation strategies.

\subsubsection{Supervised machine learning baselines}
For all of our machine learning experiments, we choose a variation of the widely successful U-Net architecture as our network \cite{Ronneberger.2015}. We follow the nnU-Net guidelines proposed by Isensee \textit{et al.} to select the image preprocessing steps, network architecture, and network training strategy \cite{Isensee.2021}. The nnU-Net training strategy has been found to be successful on a variety of segmentation tasks and is considered the de facto state-of-the-art for semantic image segmentation \cite{ Isensee.2021, Gonzalez.2023, HoussamElHariri.2022}.

We first train a \textbf{supervised} baseline using the provided labels for each dataset. We apply a default augmentation in the form of random rotations by $k \cdot \ang{90} \pm \ang{10}, k\in\{0,1,2,3\}$ and random flipping. The network is trained for 60 epochs with a batch size of 4. 

\subsubsection{Traditional computer vision baselines}
\textbf{Frangi} filters are a commonly used method to enhance the visibility of tubular structures in images. We follow the recommendations by Giarratano \textit{et al.} and select the filter scales $\sigma \in [0.5, 1, 1.5, 2]$,  the Frangi correction constant for plate-like structures $\alpha=1$ and for blob-like structures $\beta=15$. The enhanced image is binarized using thresholding while small objects are removed from the segmentation map. We adjust the threshold to align the prediction with the label map of each dataset. We use the Bayesian Optimization HyperBand (BOHB) search algorithm with 20 concurrent actors and five iterations to find the optimal threshold and minimal object size for each training set \cite{Falkner.2018}.

We also test a 2D implementation of the optimal oriented flux (\textbf{OOF}) filter, which seems to be especially well suited for dense segmentation tasks \cite{Law.2008}. We again use BOHB to find the optimal threshold and minimal object size. 

\subsubsection{Training with synthetic OCTA images by Menten \textit{et al.}}
We train a neural network on 500 synthetic OCTA images generated using the method by \textbf{Menten \textit{et al.}} Their dataset only offers a single level of detail, which means that the label maps include all visible vessels from the image. As described in section \ref{sec:alignment}, this does not align well with the annotation scheme of the validation datasets. We therefore also apply the label alignment step via filtering for better validation performance and to ensure a fair comparison to our method. Furthermore, we apply random rotation and flipping as augmentations and train for 30 epochs with batch size 4. To prevent any performance drops on the dataset by Giarratano \textit{et al.} caused by differences of image size during testing, we crop the training samples from the models trained on synthetic images to the same number of pixels.

\subsubsection{Training with our synthetic OCTA images}
We now describe the training procedure for our proposed methods. We first train a baseline model using our synthetic images without any of the proposed augmentation strategies (\textbf{S}). The network is trained on 500 synthetic images, and we use the same training procedure as for training on the images by Menten \textit{et al.}
In a second setup, we augment the training samples using our proposed handcrafted noise model with random control point values within a heuristically chosen interval (\textbf{S+RNM}, i.e., random noise model). We set $\lambda_\Delta=1$, $\lambda_N=0.7$, and $\lambda_\Gamma=0.3$.
Next, we test our adversarial training approach to optimize the control points (\textbf{S+ANM}, i.e., adversarial noise model). We set the adversarial learning rate to $lr=1e^{-3}$ and use three adversarial optimization steps.
Finally, we test our image noise model based on GAN augmentation (\textbf{S+GAN}). We first train the generator network with complete label maps for 50 epochs and then use the trained generator to augment synthetic images. While the trained segmentation network can be used for evaluation directly, we decide to train a separate nnU-Net on the transformed images with the exact same setting as in the other experiments to ensure a fair comparison. We also experience more stable training of the GAN when including all vessels vs. a filtered vessel label map (as needed for OCTA-500 and ROSE-1).

\subsection{Evaluation}
We perform 5-fold cross-validation for all methods. For Frangi and OOF, we use the training set of each fold to find the best threshold and minimal object size. For U-Nets trained on synthetic data, we train five separate models and validate them on the test split. The obtained predictions are quantitatively compared to the provided annotations using six different metrics. We calculate the mean Dice similarity coefficient (DSC), centerlineDice (clDice) \cite{Shit.2021}, area under the receiver operating characteristic curve (AUC), accuracy, recall, and precision with their respective standard errors.
However, we note that while the quantitative comparison does provide some insights into a network's predictions, the results need to be taken with a grain of salt. The provided annotations are often imperfect and therefore do not allow to accurately assess the quality of more detailed segmentation maps obtained by the automated methods.

\section{Results}
\subsection{Simulation of the retinal vasculature}
Figure \ref{figure:synth-comparison} shows a comparison of synthetic images generated by Menten \textit{et al.}, real images, and our method. The new framework allows intuitive tuning of parameters and straightforward formulation of additional constraints to create more realistic images. GAN augmentation further reduces the domain gap between synthetic images and the target dataset. Using a Python implementation with k-d trees for efficient queries based on coordinates, we are able to generate images with more than 10,000 nodes in less than a single minute, compared to the eight hours required by Menten \textit{et al.}'s simulation. The reduced runtime also simplifies the testing of new configurations, as their effects can be explored quickly. This previously hindered the practical adaptation of the model by Menten \textit{et al.} The new python implementation of our algorithm can more easily be adapted to include additional constraints for modeling specific disease spectra.

\subsection{Vessel segmentation performance}
\begin{figure}[htbp]
  \centering
  \includegraphics[width=0.79\columnwidth]{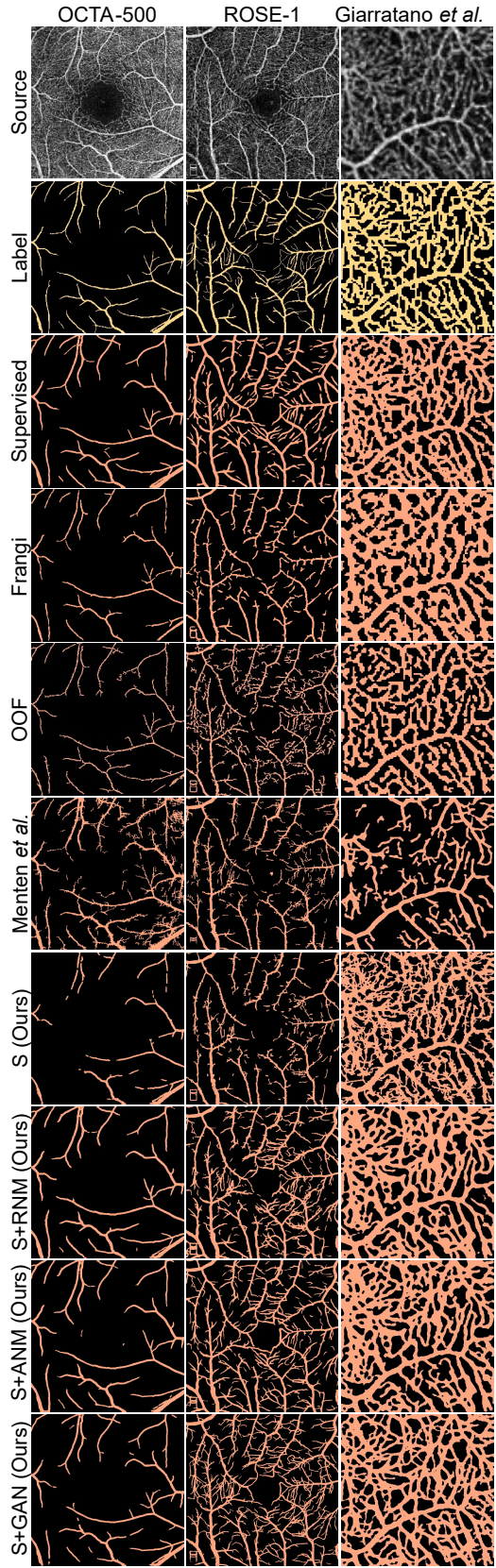}
  \caption{Qualitative comparison of all eight tested segmentation methods on a representative example from each of the three considered datasets. Our methods with additional contrast adaptations are reliable in extracting realistic segmentation maps across all datasets.}
  \label{figure:qual0}
\end{figure}

\begin{figure}[htbp]
  \centering
  \includegraphics[width=0.9\columnwidth]{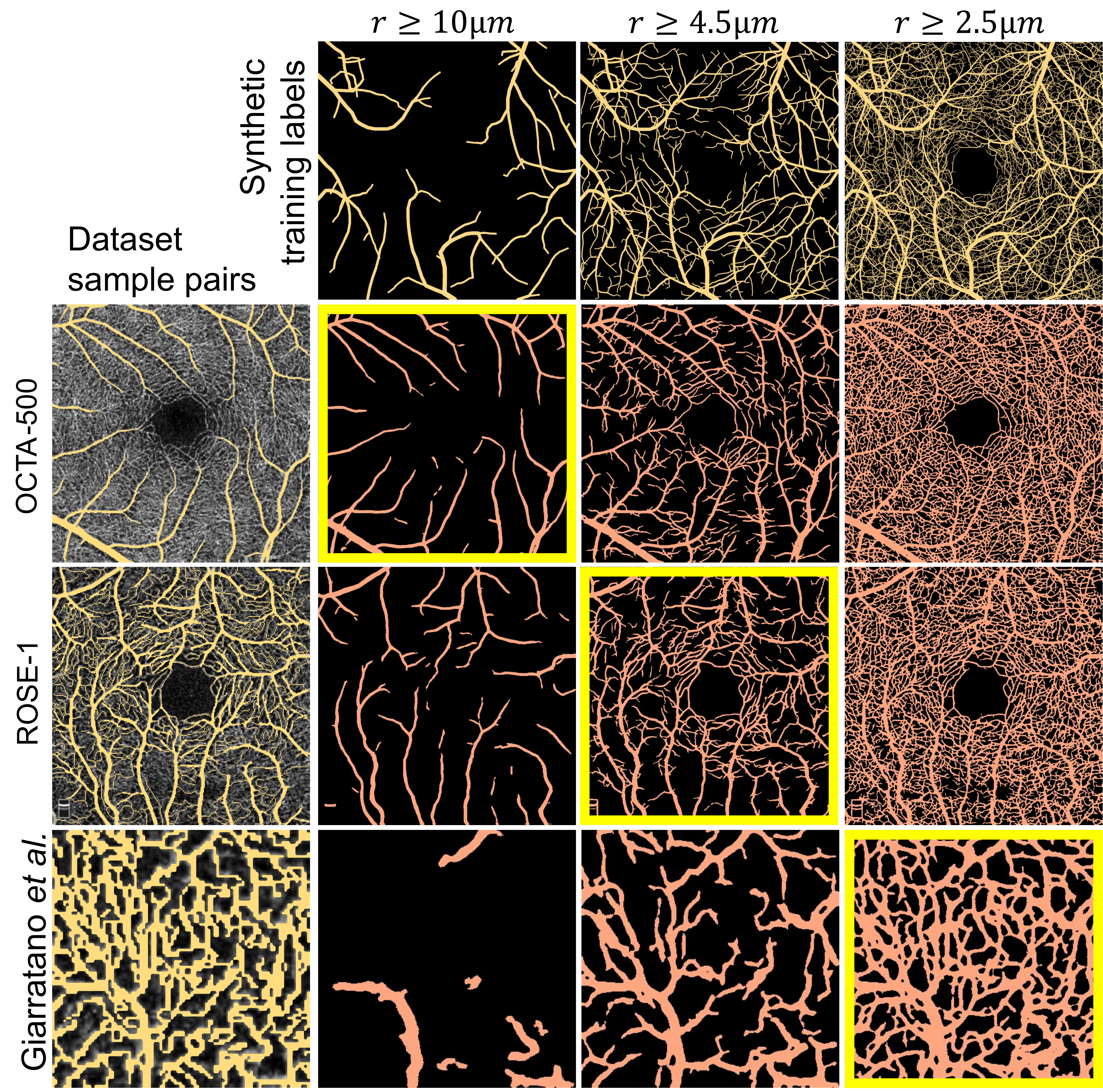}
  \caption{By controlling the minimal vessel radius in the synthetic vessel annotations, the network can be conditioned to create segmentations with varying levels of detail. This enables benchmarks on datasets with different label predictions.}
  \label{figure:qual1}
\end{figure}

\begin{figure}[htbp]
  \centering
  \includegraphics[width=\columnwidth]{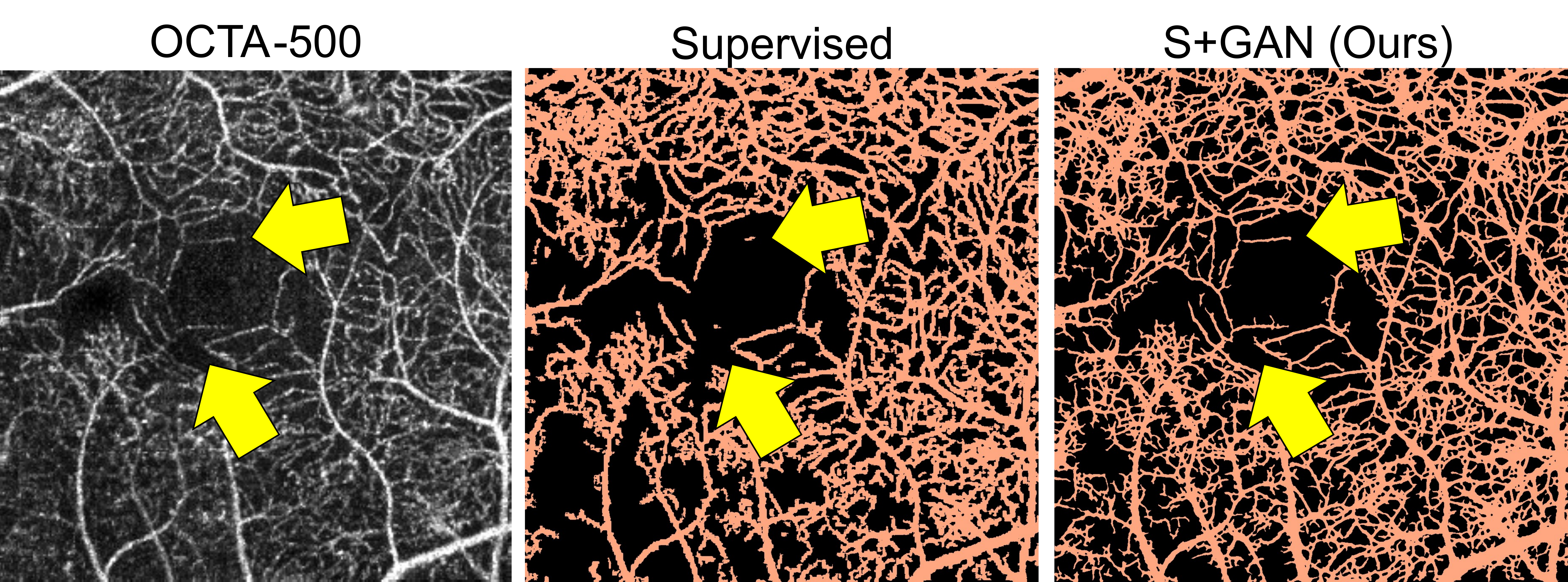}
  \caption{Performance of our and the supervised method on a DR patient. Our approach is superior in extracting detailed vessel maps in low contrast areas, such as around the FAZ.}
  \label{figure:qual}
\end{figure}

\begin{figure}[htbp]
  \centering
  \includegraphics[width=\columnwidth]{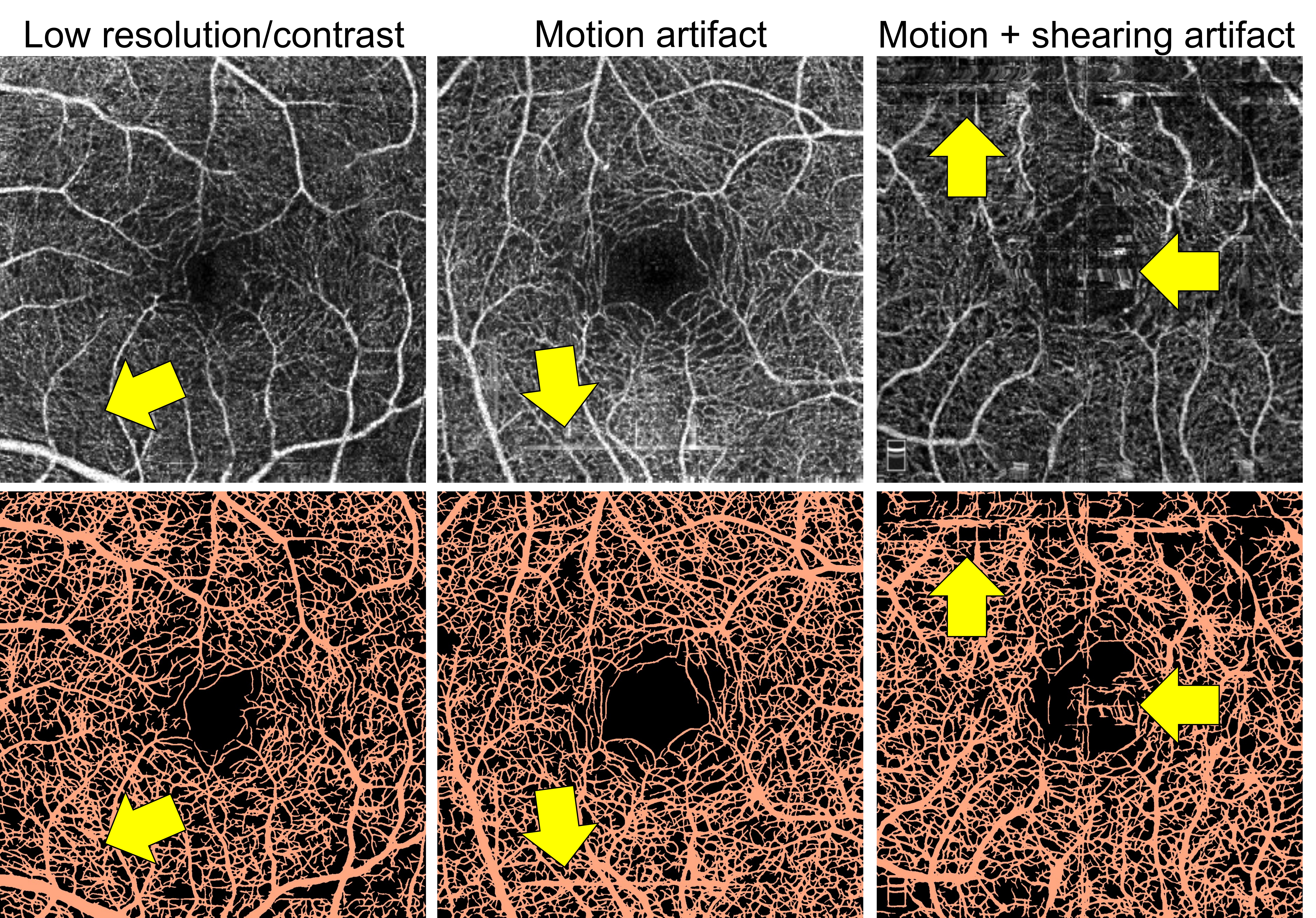}
  \caption{Our method is able to recover vessels in low resolution areas that would be difficult to annotate manually. The network interprets stronger motion and shearing artifacts as vessels, since we did not explicitly steel our network against them.}
  \label{figure:artifacts}
\end{figure}

\begin{table*}[htbp]
\caption[]{5-fold cross-validation segmentation performance of all methods on the three evaluation datasets.
The best (underline + bold) and the second best (bold) results are marked. Our methods outperform traditional computer vision algorithms. }\label{tab:results}
\begin{adjustbox}{center}
\resizebox{\textwidth}{!}{
\begin{tabular}{|l|c|c|c||cccccc|}
 \hline
 \TstrutL
 \textbf{OCTA-500}&
  \textbf{Model} &
  \textbf{Trained on} &
  \textbf{Augmentation} &
  \textbf{DSC} &
  \textbf{clDice} &
  \textbf{AUC} &
  \textbf{ACC} &
  \textbf{Recall} &
  \textbf{Precision} \\[1.5pt] \cline{1-10}
  
  \Tstrut
  Supervised &
  nnU-Net &
  Real &
  Default &
  \underline{\textbf{0.912$\pm$0.001}} &
  \underline{\textbf{0.940$\pm$0.002}} &
  \underline{\textbf{0.950$\pm$0.001}} &
  \underline{\textbf{0.984$\pm$0.000}} &
  \underline{\textbf{0.910$\pm$0.003}} &
  \underline{\textbf{0.916$\pm$0.003}} \\[.5pt] \cline{1-10}
  
 \Tstrut
\multirow{2}{*}{Traditional} &
  Frangi &
  \multirow{2}{*}{-} &
  \multirow{2}{*}{-} &
  0.807$\pm$0.003 &
  0.848$\pm$0.004 &
  0.895$\pm$0.002 &
  \textbf{0.975$\pm$0.000} &
  0.802$\pm$0.005 &
  0.820$\pm$0.003 \\ 
  
  &
  OOF &
  &
  &
  0.734$\pm$0.004 &
  0.785$\pm$0.003 &
  0.851$\pm$0.004 &
  0.966$\pm$0.001 &
  0.719$\pm$0.009 &
  0.760$\pm$0.006 \\[.5pt] \cline{1-10}

 \Tstrut
Menten \textit{et al.} &
  nnU-Net
  &
  Synthetic &
  Menten \textit{et al.} &
  0.765$\pm$0.006 &
  0.779$\pm$0.009 &
  0.889$\pm$0.003 &
  0.954$\pm$0.002 &
  0.809$\pm$0.009 &
  0.732$\pm$0.017 \\[.5pt] \cline{1-10}
  
 \Tstrut
\multirow{4}{*}{Ours} &
  \multirow{4}{*}{nnU-Net} &
  \multirow{4}{*}{Synthetic} &
  Default &
  0.734$\pm$0.006 &
  0.731$\pm$0.009 &
  0.827$\pm$0.005 &
  0.968$\pm$0.001 &
  0.663$\pm$0.010 &
  0.827$\pm$0.003 \\
  
  &
  &
  &
  RNM &
  0.841$\pm$0.002 &
  0.880$\pm$0.003 &
  0.897$\pm$0.002 &
  0.972$\pm$0.000 &
  0.805$\pm$0.005 &
  0.884$\pm$0.004 \\

  &
  &
  &
  ANM &
  0.842$\pm$0.002 &
  0.883$\pm$0.003 &
  0.898$\pm$0.003 &
  0.972$\pm$0.002 &
  0.806$\pm$0.007 &
  \textbf{0.886$\pm$0.004}\\
  
  &
  &
  &
  GAN &
  \textbf{0.866$\pm$0.001} &
  \textbf{0.916$\pm$0.002} &
  \textbf{0.940$\pm$0.002} &
  \textbf{0.975$\pm$0.000} &
  \textbf{0.898$\pm$0.003} &
  0.839$\pm$0.013\\ \hline
  
   \multicolumn{1}{c}{}\\

 \hline
 \TstrutL
 \textbf{ROSE-1}&
    \textbf{Model} &
    \textbf{Trained on} &
    \textbf{Augmentation} &
    \textbf{DSC} &
    \textbf{clDice} &
    \textbf{AUC} &
    \textbf{ACC} &
    \textbf{Recall} &
    \textbf{Precision} \\[1.5pt] \hline
  
 \Tstrut
Supervised &
    nnU-Net &
    Real &
    Default &
    \underline{\textbf{0.717$\pm$0.004}} &
    \underline{\textbf{0.700$\pm$0.006}} &
    \underline{\textbf{0.799$\pm$0.003}} &
    0.843$\pm$0.003 &
    \underline{\textbf{0.678$\pm$0.006}} &
    \underline{\textbf{0.774$\pm$0.005}}
    \\[.5pt] \cline{1-10}
    
   \Tstrut
\multirow{2}{*}{Traditional} &
    Frangi &
    \multirow{2}{*}{-} &
    \multirow{2}{*}{-} &
    \textbf{0.670$\pm$0.003} &
    0.601$\pm$0.007 &
    \textbf{0.789$\pm$0.003} &
    \underline{\textbf{0.875$\pm$0.002}} &
    \textbf{0.644$\pm$0.009} &
    0.704$\pm$0.012 \\
    
    &
    OOF &
    &
    &
    0.594$\pm$0.003 &
    0.559$\pm$0.003 &
    0.752$\pm$0.004 &
    0.831$\pm$0.006 &
    0.623$\pm$0.016 &
  0.578$\pm$0.018 \\[.5pt] \hline
  
 \Tstrut
Menten \textit{et al.} &
    nnU-Net &
    Synthetic &
    Menten \textit{et al.} &
    0.591$\pm$0.005 &
    0.615$\pm$0.005 &
    0.713$\pm$0.003 &
    0.798$\pm$0.004 &
    0.499$\pm$0.013 &
    0.720$\pm$0.004 \\ \hline
    
   \Tstrut
\multirow{4}{*}{Ours} &
    \multirow{4}{*}{nnU-Net} &
    \multirow{4}{*}{Synthetic} &
    Default &
    0.613$\pm$0.004 &
    0.546$\pm$0.004 &
    0.743$\pm$0.002 &
    \textbf{0.870$\pm$0.002} &
    0.532$\pm$0.003 &
    \textbf{0.731$\pm$0.006} \\
    &
    &
    &
    RNM &
    0.650$\pm$0.003 &
    0.656$\pm$0.005 &
    0.754$\pm$0.002 &
    0.819$\pm$0.012 &
    0.607$\pm$0.008 &
    0.715$\pm$0.010 \\

    &
    &
    &
    ANM &
    0.644$\pm$0.004 &
    0.654$\pm$0.006 &
    0.750$\pm$0.003 &
    0.809$\pm$0.004 &
    0.593$\pm$0.013 &
    0.721$\pm$0.014\\

    &
    &
    &
    GAN &
    \textbf{0.670$\pm$0.005} &
    \textbf{0.666$\pm$0.006} &
    0.768$\pm$0.003 &
    0.818$\pm$0.004 &
    0.638$\pm$0.008 &
    0.720$\pm$0.007\\ \hline

    \multicolumn{1}{c}{}\\

\hline
\TstrutL
  \textbf{Giarratano \textit{et al.}} &
    \textbf{Model} &
    \textbf{Trained on} &
    \textbf{Augmentation} &
    \textbf{DSC} &
    \textbf{clDice} &
    \textbf{AUC} &
    \textbf{ACC} &
    \textbf{Recall} &
    \textbf{Precision} \\[1.5pt] \cline{1-10}\hline
    
\Tstrut
Supervised &
    nnU-Net &
    Real &
    Default &
    \underline{\textbf{0.907$\pm$0.002}} &
    \underline{\textbf{0.954$\pm$0.003}} &
    \underline{\textbf{0.887$\pm$0.004}} &
    \underline{\textbf{0.895$\pm$0.002}} &
    \underline{\textbf{0.925$\pm$0.003}} &
    \underline{\textbf{0.981$\pm$0.006}} \\[.5pt] \cline{1-10}

\Tstrut    
\multirow{2}{*}{Traditional} &
    Frangi &
    \multirow{2}{*}{-} &
    \multirow{2}{*}{-} &
    0.769$\pm$0.009 &
    0.833$\pm$0.007 &
    0.815$\pm$0.002 &
    0.797$\pm$0.003 &
    \textbf{0.895$\pm$0.010} &
  0.683$\pm$0.012 \\[.5pt]
    &
    OOF &
    &
    &
    0.812$\pm$0.005 &
    0.848$\pm$0.006 &
    0.851$\pm$0.003 &
    \textbf{0.854$\pm$0.002} &
    0.808$\pm$0.017 &
  0.827$\pm$0.019 \\[.5pt] \hline
  
\Tstrut
Menten \textit{et al.} &
    nnU-Net &
    Synthetic &
    Menten \textit{et al.} &
    0.638$\pm$0.006 &
    0.791$\pm$0.007 &
    0.732$\pm$0.004 &
    0.687$\pm$0.010 &
    0.481$\pm$0.007 &
    \textbf{ 0.968$\pm$0.003} \\[.5pt] \hline

\Tstrut
\multirow{4}{*}{Ours} &
    \multirow{4}{*}{nnU-Net} &
    \multirow{4}{*}{Synthetic} &
    Default &
    0.781$\pm$0.006 &
    0.808$\pm$0.008 &
    0.827$\pm$0.009 &
    0.812$\pm$0.003 &
    0.887$\pm$0.003 &
    0.710$\pm$0.012 \\

    &
    &
    &
    RNM &
    \textbf{0.845$\pm$0.003} &
    \textbf{0.936$\pm$0.004} &
    \textbf{0.869$\pm$0.018} &
    0.835$\pm$0.004 &
    0.799$\pm$0.007 &
    0.904$\pm$0.007 \\
    &
    &
    &
    ANM &
    0.841$\pm$0.003 &
    0.932$\pm$0.005 &
    0.848$\pm$0.003 &
    0.831$\pm$0.005 &
    0.794$\pm$0.007 &
    0.903$\pm$0.007\\
    
    &
    &
    &
    GAN &
    0.842$\pm$0.003 &
    0.929$\pm$0.003 &
    0.842$\pm$0.003 &
    0.832$\pm$0.003 &
    0.795$\pm$0.003 &
    0.901$\pm$0.004\\ \hline
\end{tabular}
}
\end{adjustbox}
\end{table*}

Table \ref{tab:results} lists the segmentation performance of all algorithms on the three OCTA datasets and figure \ref{figure:qual0} provides a qualitative comparison for each method. The supervised model performs best quantitatively on all datasets given the manually annotated labels. The performance matches the results reported in previous works, which supports our choice of the nnU-Net architecture \cite{Giarratano.2020,Ma.2021,Li.14.12.2020}. Notably, the performance on the ROSE-1 dataset is relatively low for all methods. The predicted label maps often contain too many or too few annotations. All models struggle to decide which vessels need to be segmented and which can be ignored. 
During the evaluation of the performance of the supervised method, it is crucial to distinguish between these quantitative metrics and the accuracy w.r.t. the true vascular anatomy. A model trained under supervision of human labels tends to mimic the annotators' labeling policy, which may not necessarily represent the correct labeling schema. We elaborate on this point in section \ref{chapter:discussion}.

Of the traditional computer vision methods, the Frangi filter performs best on larger vessel sizes, while OOF works best for dense segmentation.
The training samples from Menten \textit{et al.} only offer limited complexity in the simulated vessel graphs and inferior contrast adaptation, which prevent a detailed segmentation on the dataset by Giarratano \textit{et al.}
While the method does achieve a high recall for large vessels and a high precision on the dataset by Giarratano \textit{et al.}, it generally underperforms in all experiments with regard to the Dice score.

Training on our synthetic training samples without contrast adaptations performs similar to Frangi and OOF on all datasets. Figure \ref{figure:qual1} shows that the pipeline can be adjusted to align the level of detail with the respective dataset labels. 

Adding any of our proposed contrast adaptation strategies boosts the performance notably and enables the network to outperform traditional computer vision algorithms. We found that our noise model with random control points performs similarly well in terms of the quantitative results when compared with the adversarial and the GAN-based method on the considered OCTA datasets. Figure \ref{figure:artifacts} shows that the trained network is able to identify large and medium-sized vessels in noisy and low-contrast regions, and smaller vessels in high-contrast areas. However, it struggles with motion and shearing artifacts, as commonly encountered in the ROSE-1 dataset.

A qualitative comparison for detailed segmentation in figure \ref{figure:qual} shows that existing methods are not able to extract realistic label maps. For clinical utility it is especially important to be able to reliably segment vessels for out of distribution cases, such as patients with disease. The supervised model trained on the dataset by Giarratano \textit{et al.} does not recognize vessels with lower visibility around the FAZ. In contrast, our proposed method is more robust, even for progressed DR patients. 

We hypothesize that this discrepancy between quantitative and qualitative performance stems from the quality of the available annotations that are used as ground truth. Any deviation from the label map is penalized, even if the predicted segmentation map includes more details.

\subsection{Ablation studies}
We conduct a series of experiments to comprehensively evaluate the contributions of each building block in our pipeline.
We first investigate the design of our handcrafted noise model. Table \ref{table:ablation_noise} shows the benefits of adding each of our chosen transformations to the training images. The results support our intuition that training benefits from structured vessel noise, speckle noise, contrast variations, and blurring caused by downsampling.

\begin{table}[htbp]
\centering
\caption{Ablation study of the components in our handcrafted noise model tested on the OCTA-500 dataset. Background vessel noise $\Delta$, speckle noise $N$, contrast adaptation $\Gamma$, and decreasing the resolution $\downarrow \uparrow$ each improve the segmentation performance.}
\label{table:ablation_noise}
\begin{tabular*}{\linewidth}{@{\extracolsep{\fill}}cccc}
\hline
\Tstrut
Model & Trained on & Augmentation & DSC\\ \hline
\Tstrut
\multirow{5}{*}{nnU-Net} &
\multirow{5}{*}{Synthetic} &
- & 0.734$\pm$0.006 \\
&&$\Delta$&0.805$\pm$0.004\\
&&$\Delta + N$&0.819$\pm$0.002\\
&&$\Delta + N + \Gamma$&0.836$\pm$0.001\\
&&$\Delta + N + \Gamma + \downarrow \uparrow$ & \textbf{0.841$\pm$0.002}\\
\hline
\end{tabular*}
\end{table}

Next, we compare our contributions for vessel simulation and data augmentation alongside the method by Menten \textit{et al.} \cite{Menten.2022}. For this, we measure the downstream segmentation performance of the following configurations: First, we apply the handcrafted augmentations proposed by Menten \textit{et al.} to our synthetic vessel maps. Second, we train our GAN pipeline using the vessel maps from Menten \textit{et al.} The results presented in Table  \ref{table:ablation_menten} demonstrate the enhanced vessel segmentation performance achieved through the combined effect of both our novel vessel simulation and the image-to-image augmentation. Notably, the incorporation of small capillaries yielded the most substantial performance gains, particularly evident in the dataset by Giarratano et al. GAN training using the vessel maps by Menten \textit{et al.} proved to be less stable, attributed to a lack of realism in the geometrical structure of their data compared to ours.

\begin{table*}[htbp]
\centering
\caption{Comparison of the vessel simulation and image transformation strategy by Menten \textit{et al.} and ours. Our vessel simulation enables the modeling of smaller capillaries and leads to better performance on detailed segmentation. GAN augmentation is superior to the handcrafted approach by Menten \textit{et al.} and leads to considerable performance gains.}
\label{table:ablation_menten}
\begin{tabular*}{\linewidth}{@{\extracolsep{\fill}}ccccc}
\hline
\Tstrut
\textbf{Vessel simulation} & \textbf{Augmentation strategy} & \textbf{OCTA-500} & \textbf{ROSE-1} & \textbf{Giarratano \textit{et al.}}\\ \hline
\Tstrut
Menten & Handcrafted (Menten) & 0.765$\pm$0.006 & {0.591$\pm$0.005}& 0.638$\pm$0.006\\
Ours & Handcrafted (Menten) & 0.792$\pm$0.008 & 0.593$\pm$0.005 & 0.777$\pm$0.007\\
Menten & GAN (Ours) & 0.776$\pm$0.030 & 0.603$\pm$0.011 & 0.578$\pm0.028$ \\
Ours & GAN (Ours) & \textbf{0.866$\pm$0.001} &\textbf{0.670$\pm$0.005} & \textbf{0.842$\pm$0.003} \\
\hline
\end{tabular*}
\end{table*}

Finally, we compare the performance of our proposed GAN framework with five established image-to-image frameworks:
\begin{enumerate}
    \item CycleGAN \cite{Zhu.2017}: One of the first methods for effective unpaired image-to-image translation
    \item NICE-GAN \cite{Chen_2020_CVPR}: Popular variant of CycleGAN, where the generator reuses layers of the discriminator
    \item CUT \cite{park2020contrastive}: More lightweight framework based on patchwise contrastive learning and adversarial learning
    \item NEGCUT \cite{Wang_2021_ICCV}: Version of CUT that generates instance-wise hard negatives to optimize contrastive learning
    \item DCLGAN \cite{Han_2021_CVPR}: Framework that combines the architectures of CycleGAN and CUT
\end{enumerate}
Table \ref{table:ablation_gan} lists the downstream segmentation Dice similarity coefficient obtained from three different GAN checkpoints. The table demonstrates the stability of our proposed image-to-image translation in enhancing the realism of synthetic vessel maps across all generative frameworks. The optimal overall result was observed using our method after 50 epochs of GAN training, validating our design choice. We attribute this to the effectiveness of the segmentation loss as a cycle consistency term during training. It should be noted that this performance comes at the expense of a higher memory footprint, since we work on the upscaled segmentation masks. The CUT framework yielded strong performance using minimal GPU resources, making it a suitable alternative for budged applications. The NICE-GAN framework is an exception in the table, as it performs significantly worse than its counterparts. Manual inspection of the generated images suggests that the generator not only altered the style, but also changed the vessel structure. The resulting misalignment between vessels and labels then leads to a decrease in performance during downstream segmentation tasks. A similar effect is noted after prolonged GAN training in the other frameworks.

\begin{table}[htbp]
\centering
\caption{Comparison of different GAN frameworks for unpaired synthetic-to-realistic OCTA image translation. All methods were trained on the OCTA-500 dataset using the same training schema with the network specific default hyperparameters. The checkpoints after epoch 25, 50, and 100 were then used to train a nnU-Net for vessel segmentation. We performed 5-fold cross validation on both the GANs and the nnU-Nets.}
\label{table:ablation_gan}
\begin{tabular*}{\linewidth}{@{\extracolsep{\fill}}cccc}
\hline
\Tstrut
\textbf{Model} & \textbf{DSC 25} & \textbf{DSC 50} & \textbf{DSC 100}\\ \hline
\Tstrut
CycleGAN & \textbf{0.852$\pm$0.004} & 0.818$\pm$0.008 & \textbf{0.826$\pm$0.014}\\
NICE-GAN & 0.336$\pm$0.081 & 0.358$\pm$0.093 & 0.391$\pm$0.082\\
CUT & 0.845$\pm$0.004 & 0.833$\pm$0.004 & 0.820$\pm$0.006\\
NEGCUT & 0.836$\pm$0.005 & 0.828$\pm$0.005 & 0.819$\pm$0.003\\
DCLGAN & 0.843$\pm$0.006 & 0.827$\pm$0.009 & 0.837$\pm$0.005\\
Ours & 0.836$\pm$0.004 & \textbf{0.866$\pm$0.001} & \textbf{0.826$\pm$0.004}\\
\hline
\end{tabular*}
\end{table}

\section{Discussion and Conclusion}
\label{chapter:discussion}
In this work, we proposed a method to generate highly realistic synthetic OCTA images to train a CNN for vessel segmentation in real images. We substantially surpass the realism of the generated vascular networks proposed by Menten \textit{et al.} Our new simulation based on space colonization creates more detailed vascular graphs in less than one minute. We proposed three contrast adaptation strategies to minimize the domain gap between synthetic and real images, with the goal of improving the robustness of the trained network. Using these, we are able to extract detailed vessel segmentation maps from OCTA \textit{en-face} images without any human annotations. In experiments on three public datasets, we outperform established computer vision algorithms.

Differences in the degree of detail for provided annotations in existing validation datasets require aligning the level of segmentation detail of the training labels with those of the respective validation dataset. To this end, we proposed to filter vessels in the label map by diameter. However, the manual annotations do not always correlate with vessel diameter and inconsistencies in labeling impede the alignment of synthetic annotations. Poorly visible vessels, even if they were relatively large in diameter, were often ignored by the annotators. Therefore, quantitative comparisons with existing datasets labels do not allow to accurately assess the quality of detailed segmentation maps. Any deviations from the human annotations, even if correct, are penalized.

Because of this limited expressiveness, we draw qualitative comparisons on dense segmentation for unseen data.
While the OOF filter achieves satisfying performance for areas with good image quality, it struggles to connect vessel segments in low-contrast regions.
The low amount of training data limits the performance of the supervised model to recognize vessels with low visibility. This becomes especially pronounced around the FAZ and in progressed DR patients.
In contrast, our approach is especially well suited to extract realistically detailed segmentations and surpasses every other method on unseen samples from the OCTA-500 dataset. Supervised methods are bound to the quality of human annotations, which often under- or overestimate the vessel diameter in low resolution or low contrast areas. In contrast, our method automatically learns a robust representation of how vessels look like and how the seen footprint relates to the true vessel diameter.

Future work should concentrate on exploring the usefulness of extracted biomarkers for downstream tasks. Taking the example of DR classification, one could compare the performance between direct classification from the image, or based on classification using extracted biomarkers. By taking the extra step of determining biomarkers, we provide clinicians with more interpretability of the machine learning pipeline. It is also possible to extend our framework to 3D OCTA volumes by adjusting the simulation model to realistically recreate the retinal anatomy in 3D. By controlling the placement of the OSs and vessel growth hyperparameters in the simulation model, it is possible to simulate different vascular layers. This would provide segmentation maps with additional depth information, enabling more accurate analysis.

We make our entire pipeline, pretrained models, as well as a large dataset of synthetic OCTA images publicly available at \url{https://github.com/TUM-AIMED/OCTA-seg}. Our tool allows the automated extraction of detailed segmentation maps and enables the computation of quantitative biomarkers. By this, we hope to make a valuable contribution to the ophthalmology community and motivate more research in automated OCTA analysis.

\bibliographystyle{IEEEtran}
\bibliography{main}

\end{document}